\documentclass[twocolumn,showpacs,nofootinbib,amsmath,amssymb,prd]{revtex4}
\usepackage{hyperref}
\usepackage{amsfonts}
\usepackage{graphicx}

\begin{document}

\def\covder{\nabla}
\def\lapse{\alpha}
\def\psiana{\rho}
\newcommand{\R}{\mathrm{I}\kern -2.2pt \mathrm{R}}

\title{Can a combination of the conformal thin-sandwich and puncture methods 
yield binary black hole solutions in quasi-equilibrium?}

\author{Mark D. Hannam}
\email{marko@physics.unc.edu}
\affiliation{Department of Physics and Astronomy, University of North 
Carolina, Chapel Hill, North Carolina 27599}
\author{Gregory B. Cook}
\email{cookgb@wfu.edu}
\affiliation{Theoretical Astrophysics 130-33, California Institute of 
Technology, Pasadena, California 91125}
\altaffiliation[Permanent address: ]{Department of Physics, Wake Forest 
University, Winston-Salem, North Carolina 27109}
\author{Thomas W. Baumgarte}
\email{tbaumgar@bowdoin.edu}
\affiliation{Department of Physics and Astronomy, Bowdoin College,
Brunswick, Maine 04011}
\altaffiliation[Also at: ]{Department of Physics, University of 
Illinois at Urbana-Champaign, Urbana, Illinois 61801} 
\author{Charles R. Evans}
\email{evans@physics.unc.edu}
\affiliation{Department of Physics and Astronomy, University of North 
Carolina, Chapel Hill, North Carolina 27599}

\begin{abstract}
We consider combining two important methods for constructing
quasi-equilibrium initial data for binary black holes: the conformal
thin-sandwich formalism and the puncture method.  The former seeks to
enforce stationarity in the conformal three-metric and the latter
attempts to avoid internal boundaries, like minimal surfaces or
apparent horizons.  We show that these two methods make partially
conflicting requirements on the boundary conditions that determine the
time slices.  In particular, it does not seem possible to construct
slices that are quasi-stationary and avoid physical singularities and
simultaneously are connected by an everywhere positive lapse function,
a condition which must obtain if internal boundaries are to be avoided.  
Some relaxation of these conflicting requirements may yield a soluble
system, but some of the advantages that were sought in combining these
approaches will be lost.
\end{abstract}

\pacs{04.20.Ex, 04.25.Dm, 04.30.Db, 04.70.Bw}

\maketitle

\section{Introduction}

Binary black holes are expected to produce some of the strongest 
gravitational wave signals that lie within the pass-bands of both the new 
generation of terrestrial gravitational wave observatories, such as LIGO, 
GEO, TAMA, and VIRGO, and the space-based LISA detector.  All aspects of
the orbital decay and signal history of such decaying binaries are of
interest, both for enhancing the prospects of source detection and for
determining source parameters~\cite{thorne95,cf94}.  With this need 
in mind, an intense worldwide theoretical effort is being made to model 
different phases of the merger of binary black holes.  

Considerable interest attaches to the phase between early inspiral and 
loss of orbital stability and plunge.  For astrophysical reasons many binary
black holes are expected to enter this phase in quasi-circular orbital 
motion, with the residual orbital eccentricity following merely from the 
gradual radiative loss of energy and angular momentum.  Hence, early on 
these quasi-circular orbits are well approximated by positing circular 
motion and ignoring radiation reaction.

The first numerical models of binary black holes in quasi-circular
orbit were obtained by Cook \cite{cook94} (hereafter C94), who used
an effective-potential method to identify those particular data sets
that most closely approximated binaries in circular orbits.  The
initial data were constructed using the conformal-imaging formalism.
In that approach, the momentum constraints decouple from the
Hamiltonian constraint and can be solved analytically, yielding the
Bowen-York conformal extrinsic curvature \cite{bowen80}.  The
conformal-imaging approach takes as given a conformally-flat metric, a 
maximal time slice, and a two-sheeted topology where two identical 
universes are connected by one or more black holes.  The identification of 
the universes manifests as an isometry condition that restricts the form
of the solutions.  This isometry condition can be used to obtain
boundary conditions on the black-hole throats.  In particular, a
minimal surface boundary condition is obtained for the conformal
factor and the Hamiltonian constraint need only be solved on one of the
two isometric sheets of the hypersurface.  The solution of the Hamiltonian 
constraint is (near completely) restricted to the exterior of the black 
holes in one of the two universes and the calculation avoids both 
coordinate and physical singularities in the black hole interiors.

Brandt and Br\"ugmann \cite{brandt97} suggested an alternative
{\sl puncture method} for the construction of spacetimes containing 
multiple black holes.  The momentum constraint is again solved 
analytically on the conformal manifold, yielding the Bowen-York extrinsic 
curvature.  The conformal factor, however, is split into a sum of an 
analytically-known singular term and a correction.  The analytic term 
corresponds to the conformal factor of superposed static black holes in 
isotropic coordinates and captures the topological character of the black 
holes.  The Hamiltonian constraint is then written as an equation for the
correction term, which exists for non-vanishing extrinsic curvature.  In 
this method, though, isometry across the throats is not assumed or used.  
Instead the computational domain is all of $\R^3$ minus the puncture points 
where the conformal factor is singular.  However, by construction, the 
terms in the Hamiltonian constraint are regular everywhere, allowing the 
punctures to be ignored.  By eliminating the need to excise the black hole 
interiors and to impose boundary conditions on the black hole throats, the 
puncture method provides a much simpler computational domain.  
Baumgarte~\cite{baum00} (hereafter B00) adopted this approach to construct 
binary black holes in quasi-circular orbits and found very similar results 
to those of C94.  This is not surprising, as the resulting spacetimes only 
differ in terms of the underlying topology (the puncture method gives rise
to $N+1$ asymptotically flat regions with $N$ throats), and the effects 
can be estimated to be fairly small.

Gourgoulhon, Grandcl\'ement and Bonazolla \cite{ggba,ggbb} (hereafter
GGB02) adopted a conformal thin-sandwich decomposition of the initial
value problem (e.g.~\cite{york98,cook00}) to construct binary black
hole data.  Unlike the methods described above, the conformal 
thin-sandwich decomposition has a link to dynamics.  This is
seen in the appearance of the lapse function and shift vector in the
conformal thin-sandwich equations, and in the fact that the time rate
of change of the conformal three-geometry is part of the
freely-specifiable data.  In the conformal thin-sandwich approach, the
momentum constraint does not yield the Bowen-York extrinsic curvature
as a solution.  Instead, it becomes an elliptic system on the shift
vector, which once solved indirectly determines the extrinsic curvature.  
As before, the Hamiltonian constraint is an elliptic equation for the 
conformal factor but the conformal thin-sandwich approach must be 
supplemented with a condition on the lapse.  GGB02 adopted maximal slicing, 
which is a natural choice for the construction of equilibrium data.  They 
also take the time rate of change of the conformal three geometry to vanish
which again is a natural choice for equilibrium data.  GGB02 also
follow C94 in assuming an isometry across the throats, which allows
the black hole interiors to be excised.  While this approach is more
involved than those of C94 and B00, it has been suggested that the
conformal thin-sandwich decomposition is more natural for the
construction of binaries in quasi-circular orbit (e.g.~\cite{cook00}),
and in fact the results of GGB02 agree more closely with post-Newtonian
calculations than do those of C94 and B00 \cite{blanchet02,dgg02}.  There 
is furthermore good evidence that the differences between the results of 
GGB02 and those of C94 and B00 are due in fact to the differences in 
initial value decompositions, in particular the behavior of the extrinsic
curvature~\cite{pfeiffer02,dgg02,sb02}.

As we will discuss in more detail in Section \ref{sec:equi}, the
imposition of the isometry on the throats leads to a technical problem
in the conformal thin-sandwich approach.  GGB02 had to regularize
their solutions on the throats in a way that introduced a small 
inconsistency (i.e.~a small violation of the constraints) into their 
solutions.  It is thought that this problem may be circumvented with a 
different set of boundary conditions on the throats \cite{cook02}.  
This issue, however, suggests that one might try to combine the 
thin-sandwich approach with the puncture approach, with the goal of 
eliminating the need for interior boundary conditions.

We have attempted to combine the conformal thin-sandwich and puncture 
methods, but have concluded that this approach leads to some conceptional 
problems and some likely serious numerical obstacles.  As we
will argue below the approach requires several simultaneous conditions:  
(1) The construction of quasi-equilibrium data requires the existence of a 
quasi-stationary coordinate system.  (2) The puncture method requires 
physical fields in the vicinity of each puncture such that each puncture 
represents a separate spatial infinity on a multiply-connected manifold 
(i.e., the slice(s) must thread each wormhole and avoid the physical 
singularities).  (3) Finally, the conformal thin-sandwich decomposition 
(usually) requires that the lapse be everywhere positive, to avoid the
existence of internal boundaries necessitating regularity conditions.  
Unfortunately, even in Schwarzschild spacetime such a slicing cannot
exist, suggesting that the construction of black hole equilibrium data
with a combined conformal thin-sandwich and puncture method is either 
impossible or will involve numerical complications that it was initially
intended to avoid.

The remainder of the paper is organized as follows.  In Section 
\ref{sec:cts} we review briefly the conformal thin-sandwich decomposition.
Section \ref{sec:equi} shows the choice of freely-specifiable data that
represents quasi-equilibrium, giving the form of the conformal thin-sandwich 
equations that must be solved.  Section \ref{sec:punc} then briefly reviews
the original puncture method.  Based on that discussion, Section 
\ref{sec:ctsp} presents how we argue that singular terms can be factored
out of the conformal factor and lapse function in the conformal thin-sandwich
equations of Section \ref{sec:equi}, allowing those equations to be solved 
as if the manifold were free of punctures.  Freedom exists in the choice
of lapse and the remainder of this section focuses on the potential 
inconsistencies or numerical difficulties in solving the momentum constraints.
Finally, Section \ref{sec:summ} summarizes our conclusions and suggests 
some avenues for further exploration.

\section{The conformal thin-sandwich decomposition}
\label{sec:cts}

The spacetime line element can be written in the \mbox{3+1} form
\begin{equation}
ds^2 = - \lapse^2 dt^2 + \gamma_{ij} (dx^i + \beta^i dt)(dx^j + \beta^j dt),
\end{equation}
where $\lapse$ is the lapse function, $\beta^i$ the shift vector, and
$\gamma_{ij}$ the induced metric on a spatial slice $\Sigma$.  The
spatial metric $\gamma_{ij}$ and the slice's extrinsic curvature
$K_{ij}$ satisfy the Hamiltonian constraint
\begin{equation} \label{eqn:HC}
R + K^2 - K_{ij} K^{ij} = 0
\end{equation}
and the momentum constraint
\begin{equation} \label{eqn:MC}
\covder_j K^{ij} - \gamma^{ij} \covder_j K = 0
\end{equation}
on each slice $\Sigma$.  Here $\covder_i$, $R_{ij}$, and 
$R = \gamma^{ij} R_{ij}$ are the spatial covariant derivative, Ricci tensor, 
and scalar curvature associated with $\gamma_{ij}$ (we have no occasion in 
this paper to use the analogous four-dimensional quantities).  We have 
assumed vanishing matter sources ($T_{\mu\nu} = 0$).  The metric and 
extrinsic curvature evolve in accordance with
\begin{equation} \label{eqn:gdot}
\partial_t \gamma_{ij} = - 2 \lapse K_{ij} 
+ \covder_i \beta_j + \covder_j \beta_i
\end{equation}
and
\begin{eqnarray} \label{eqn:Kdot}
\partial_t K_{ij} & = & - \covder_i \covder_j \lapse + \lapse (R_{ij} -
2 K_{ik} K^k_{~j} + K K_{ij}) \nonumber \\
& & + \beta^k \covder_k K_{ij} + K_{ik} \covder_j \beta^k + 
K_{kj} \covder_i \beta^k .
\end{eqnarray}

The construction of initial-data solutions requires specifying the
spatial metric and extrinsic curvature that satisfy the constraint
equations (\ref{eqn:HC}) and (\ref{eqn:MC}).  The approach based on
conformal decomposition~\cite{york79} provides a straightforward process
to solving these coupled equations.  The spatial metric
is conformally scaled by
\begin{equation}
\gamma_{ij} \equiv \psi^4 \tilde\gamma_{ij}
\end{equation}
(see \cite{lich44,york71}).  Here $\psi$ is the conformal factor and
$\tilde \gamma_{ij}$ is the conformal or background metric.  The
extrinsic curvature $K_{ij}$ is split into its trace $K$ and the
traceless part $A_{ij}$, and the latter is then also conformally 
transformed
\begin{equation}
K_{ij} = A_{ij} + \frac{1}{3} \gamma_{ij} K = \psi^{-2} \tilde A_{ij}
+ \frac{1}{3} \psi^4 \tilde \gamma_{ij} K.
\end{equation}
There is no conformal scaling of $K$.  The Hamiltonian constraint 
(\ref{eqn:HC}) can now be written as
\begin{equation}
\tilde{\covder}^2 \psi = \frac{1}{8} \psi
\tilde{R} + \frac{1}{12} \psi^5 K^2 - \frac{1}{8} \psi^{-7}
\tilde{A}_{ij} \tilde{A}^{ij}, 
\label{eqn:TSHCg} 
\end{equation}
where $\tilde \covder$, $\tilde{R}_{ij}$, and 
$\tilde{R}=\tilde{\gamma}^{ij} \tilde{R}_{ij}$ denote the covariant 
derivative, conformal Ricci tensor, and curvature scalar compatible with
the conformal metric $\tilde \gamma_{ij}$.

In the conformal thin-sandwich decomposition, both the background
metric $\tilde \gamma_{ij}$, its time derivative
\begin{equation}
\tilde u_{ij} \equiv \partial_t \tilde \gamma_{ij},
\end{equation}
and the trace of extrinsic curvature $K$ are considered freely-specifiable 
quantities.  The traceless part of the evolution equation (\ref{eqn:gdot}) 
for $\gamma_{ij}$ then yields an equation for $\tilde{A}_{ij}$
\begin{equation} 
\tilde{A}_{ij} = \frac{1}{2\tilde{\lapse}} 
\left[ (\tilde{\mathbb L} \beta)_{ij}
- \tilde{u}_{ij} \right], \label{eqn:Aeqng}
\end{equation} 
where $\tilde{\mathbb L}$ is a longitudinal operator whose action 
$(\tilde {\mathbb L} \beta)^{ij}$ on the shift yields the symmetrized 
trace-free gradient,
\begin{equation}
(\tilde{\mathbb L} \beta)^{ij} = \tilde{\covder}^i \beta^j + \tilde{\covder}^j
\beta^i - \frac{2}{3} \tilde{\gamma}^{ij} \tilde{\covder}_k \beta^k,
\end{equation} 
and where we have introduced the densitized lapse (or slicing function)
\begin{equation}
\tilde \lapse = \psi^{-6} \lapse.
\end{equation}
Inserting (\ref{eqn:Aeqng}) into the momentum constraint (\ref{eqn:MC})
yields 
\begin{equation}
\tilde{\Delta}_{\mathbb L} \beta^i - (
{\tilde{\mathbb L}} \beta )^{ij} \tilde{\covder}_j \ln \tilde{\lapse} = 
\frac{4}{3} \tilde{\lapse} \psi^6 \tilde{\covder}^i K + \tilde{\lapse}
\tilde{\covder}_j \left( \frac{1}{\tilde{\lapse}} \tilde{u}_{ij} \right),
\label{eqn:TSMCg} 
\end{equation}
which can be regarded as an elliptic system for the shift vector $\beta^i$.
Here the operator $\tilde{\Delta}_{\mathbb L}$ (a vector Laplacian) is the 
divergence of the longitudinal operator $\tilde{\mathbb L}$, with the
effect
\begin{equation}
\tilde{\Delta}_{\mathbb L} \beta^i \equiv \tilde{\covder}_j (
\tilde{\mathbb L} \beta)^{ij} = \tilde{\covder}^2 \beta^i + \frac{1}{3}
\tilde{\covder}^i ( \tilde{\covder}_k \beta^k ) + \tilde{R}^i_j \beta^j.
\end{equation}
The shift vector components are conformally invariant 
($\beta^i = \tilde{\beta}^i$).  Equations (\ref{eqn:TSHCg}) and 
(\ref{eqn:TSMCg}) are incomplete in that we have yet to specify how the lapse 
$\lapse$ is chosen.  The condition on the lapse is intertwined with the 
choice of free data that yield a quasi-equilibrium state, and we take up 
these issues next.

\section{Quasi-equilibrium data in the thin-sandwich approach}
\label{sec:equi}

The construction of binary black holes in quasi-circular orbit
assumes that the emission of gravitational radiation is negligible, so
that the spacetime is in quasi-equilibrium when viewed from a
co-rotating reference frame.  The time coordinate of this reference
frame is an approximate helical Killing vector~\cite{bgm97}.

Our goal is therefore to construct initial data which, when evolved
in such a co-rotating reference frame, would lead to metric components
that are independent of time.  The conformal thin-sandwich
decomposition provides a natural framework for the construction of
quasi-equilibrium data, since we can explicitly set the time derivative
of the conformal metric to zero
\begin{equation}
\tilde u_{ij} = 0.
\end{equation}
We follow C94, B00 and GGB02 and assume conformal flatness, i.e.
\begin{equation}
\tilde \gamma_{ij} = f_{ij},
\end{equation}
where $f_{ij}$ is a flat metric.  It is furthermore reasonable
to assume that the time-derivative of the trace of the extrinsic
curvature vanishes, $\partial_t K = 0$.  With this assumption, the trace of
(\ref{eqn:Kdot}) combined with (\ref{eqn:HC}) yields
\begin{equation}
\tilde{\nabla}^2 (\psi^7 \tilde{\lapse} ) = \tilde{\lapse} \psi^7 \left(
\frac{7}{8} \psi^{-8} \tilde{A}_{ij} \tilde{A}^{ij} + \frac{5}{12}
\psi^4 K^2 \right) + \psi^5 \beta^l \tilde{\nabla}_l K .
\label{eqn:TSconstK} 
\end{equation}
In this equation, $\tilde A^{ij}$ is now given by
\begin{equation} \label{eqn:Aeqn}
\tilde{A}^{ij} = \frac{1}{2\tilde{\lapse}} (\tilde{\mathbb L} \beta)^{ij}.
\end{equation} 
We could further follow GGB02 and choose maximal slicing, $K=0$, but
the argument is more general and we merely assume that a sufficiently 
smooth choice for $K$ is made that is consistent with asymptotic flatness
and spatial slices.

Our choices of $\tilde \gamma_{ij} = f_{ij}$, $\tilde u_{ij} = 0$ and $K$
completely determine all freely-specifiable variables.  With these
choices, the Hamiltonian constraint (\ref{eqn:TSHCg}) reduces to
\begin{equation}
\tilde{\nabla}^2 \psi = - \frac{1}{8} \psi^{-7} \tilde{A}_{ij} \tilde{A}^{ij} 
+ \frac{1}{12} \psi^5 K^2 \label{eqn:TSHC}
\end{equation}
and the momentum constraint (\ref{eqn:TSMCg}) reduces to
\begin{equation}
\tilde{\lapse}\tilde{\nabla}_j \left[\frac{1}{\tilde{\lapse}} 
(\tilde{{\mathbb L}} \beta )^{ij} \right] =
\tilde{\Delta}_{\mathbb L} \beta^i - (\tilde{{\mathbb L}} \beta )^{ij} 
\tilde{\nabla}_j \ln \tilde{\lapse} = 
\frac{4}{3} \tilde{\lapse} \psi^6 \tilde{\nabla}^i K. 
\label{eqn:TSMC} 
\end{equation}
Equations (\ref{eqn:TSconstK}) through (\ref{eqn:TSMC}) together with
appropriate boundary conditions now determine the conformal factor
$\psi$, the lapse $\lapse$ and the shift $\beta^i$.  This set of
equations was derived independently by
\cite{isenberg78,wilsonmathews95} and has been used in several
applications, in particular for the construction of binary
neutron stars (see \cite{cook00,bs03} for recent reviews).

The choice of $\tilde u_{ij} = 0$ and $\partial_t K = 0$ guarantees a
limited sense of stationarity.  However, there is no assurance that
the other components of the extrinsic curvature or the conformal
factor have a vanishing first time derivative.  (Were this the case,
it would imply exact equilibrium.)  Furthermore, there is no guarantee
that these time derivatives are even small, in some appropriate norm.
There are a couple of issues to consider.

Firstly, the boundary conditions on the problem must be made consistent with 
the limited stationarity conditions given above.  Boundary conditions can
be imposed in different ways (e.g., in the puncture method via free 
parameters at the punctures--see below) but must be reflective of two 
black holes in instantaneous transverse motion or vanishing motion in a 
rotating frame.  Even then, however, parameter freedom exists and only 
some parameter choices will lead to black holes in instantaneous force 
balance as seen in the rotating frame.  Most parameter choices will imply 
an acceleration of the holes away from or toward each other, which will be 
reflected in the time derivative of $\tilde{A}_{ij}$.  All methods to date
involve a parameter search and some equilibrium criterion, such as an 
effective potential minimization or a virial theorem argument.  

Secondly, given a solution for $\psi$, $\lapse$ and $\beta^i$ following such
a parameter search, one could {\it a posteriori \/} compute the
right hand side of (\ref{eqn:Kdot}) and the trace of (\ref{eqn:gdot})
to check whether these time derivatives are sufficiently small in an 
appropriate sense.  If they are small, then one has in fact constructed 
a quasi-equilibrium orbit.  If, however, these time derivatives are not
small, the initial data may still represent a quasi-equilibrium orbit but 
in a {\em dynamical slicing}.  Even on a stationary spacetime with a 
timelike Killing vector, an arbitrary choice of the lapse or shift would 
lead generically to apparent time-dependence of the metric.  In such a 
dynamical slicing the Killing vector representing the symmetry
(a helical Killing vector in the case of binaries) would have an unknown
functional form and it is unclear how the symmetry would be established 
or exploited.  For the construction of quasi-equilibrium orbits it therefore 
seems crucial that the solution be represented in a stationary slicing.  
The slicing condition (\ref{eqn:TSconstK}) may be consistent with 
stationarity but stationarity also depends on boundary conditions.  It is 
also an open question what advantage might arise in specifying some 
$K \ne 0$ but any such choice would also need to be consistent with 
stationarity.  

In constructing their binary black hole solutions, GGB02 employed a
method that is essentially the conformal thin-sandwich decomposition
as outlined in Section \ref{sec:cts} and this section.  They demanded 
additionally that their solutions be inversion symmetric, i.e., that the 
isometry conditions across the throats be satisfied as in C94.  One choice 
for inversion symmetry of the lapse is to require, as did GGB02, that 
$\tilde \lapse = 0$ on the throats.  This condition clearly leads to 
problems in constructing the extrinsic curvature via Eq.~(\ref{eqn:Aeqn}) 
unless $(\tilde{\mathbb L}\beta)^{ij} = 0$ on the throats as well.  This 
latter condition is not the natural isometry condition on the shift, nor is 
it the boundary condition used in GGB02.  Rather, GGB02 enforced this extra 
condition, over-determining the shift, and leading to a small inconsistency 
in their numerical solution.  It may be that these problems can be resolved 
by adopting an alternative set of boundary conditions representing black 
holes in equilibrium \cite{cook02}.

Alternatively, one might hope to avoid interior boundary conditions
altogether by combining the conformal thin-sandwich approach with an
extension of the original puncture method.  To set the stage for that 
discussion, we first review briefly the puncture method.

\section{The original puncture method}
\label{sec:punc}

The puncture method can be motivated by considering the 
Brill-Lindquist \cite{brill63} solution for multiple black holes at a 
moment of time symmetry ($K = 0, \tilde A_{ij} = 0$).  In isotropic 
coordinates $\{x^i\}$, the conformal factor can be written as
\begin{equation}
\psi = 1 + \sum_n^N \frac{{\mathcal M}_n}{2 \vert {\bf x}-{\bf x}_n \vert }
= 1 + \sum_n^N \frac{{\mathcal M}_n}{2 r_n}.
\end{equation} 
This is a solution of the Hamiltonian constraint (\ref{eqn:TSHC}) for
$N$ black holes at locations ${\bf x}_n$, each parameterized by 
a mass ${\mathcal M}_n$.  For convenience we define
\begin{equation}
\frac{1}{\psiana} \equiv \sum_n^N \frac{{\mathcal M}_n}{2 r_n}
\end{equation}
so that the above solution is simply $\psi = 1 + 1/\psiana$.  Brandt
and Br\"{u}gmann \cite{brandt97} noted that for black holes with
non-zero linear and angular momentum ($\tilde{A}_{ij} \ne 0$), the 
conformal factor can be written as
\begin{equation} \label{eqn:PuncPsi}
\psi = 1 + \frac{1}{\psiana} + u, 
\end{equation} 
where $u$ is a correction term (our definition of $u$ corresponds to 
$u - 1$ in \cite{brandt97} and B00).  Having factored out $1/\psiana$, 
the Hamiltonian constraint becomes an equation for $u$,
\begin{equation}
\nabla^2 u = - \frac{1}{8} \psi^{-7} \tilde{A}_{ij} \tilde{A}^{ij} +
\frac{1}{12} \psi^5 K^2 .
\label{eqn:PMUEQN} 
\end{equation} 
Provided the trace-free extrinsic curvature $\tilde{A}^{ij}$ diverges at 
each puncture no faster than $1/r_n^3$, the term $\psi^{-7} \tilde{A}_{ij}
\tilde{A}^{ij}$ will be everywhere finite.  A similar restriction on the 
second term on the right hand side of (\ref{eqn:PMUEQN}) (Brandt and 
Br\"{u}gmann had assumed $K = 0$) requires that $K$ vanish at least as 
fast as $r_n^3$ at each puncture.  For such choices of $\tilde{A}^{ij}$ 
and $K$, $u$ will be at least a $C^2$ function everywhere.  This allows 
the Hamiltonian constraint to be solved as if the conformal manifold were 
$\R^3$, ignoring the punctures.  In this way, there is no need 
to provide inner boundary conditions on minimal surfaces (throats) or
apparent horizons \cite{cook02,pfeiffer02}.  This greatly simplifies
the solution procedure.  B00 applied the puncture method with a
(modified) Bowen-York extrinsic curvature~\cite{bowen80} (the
$1/r_n^4$ terms that guarantee isometry are dropped for the reason
given above).  Baumgarte's initial value problem consisted of this
analytic solution of the momentum constraints and a numerical solution
of (\ref{eqn:PMUEQN}).  Thus he was able to redo the calculation of
C94 with the isometry assumption replaced by the puncture approach,
constructing a sequence of black-hole binaries in quasi-circular
orbit.  Quite similar physical results were obtained.

The puncture approach relies upon the spatial slice $\Sigma$ passing 
smoothly through each wormhole and reaching a separate spatial infinity 
associated with each black hole.  The particular form of the unbounded 
growth in $\psi$ (assuming $u$ is finite) as each puncture is approached 
${\bf x} \rightarrow {\bf x}_n$ is indicative of separate compactified 
asymptotic infinities.  The intent is to avoid physical singularities, 
which could be verified by computing a Riemann curvature scalar
or, as is typical, using a set of coordinate transformations (inversion 
through spheres about each puncture) and verifying that $\psi$ and 
$\tilde{A}_{ij}$ are consistent with asymptotic flatness at the punctures. 

With the success of the puncture method in yielding results similar 
to the conformal imaging approach, it seemed worth considering a 
combination of the puncture method and the conformal thin-sandwich 
formalism.  Such a method has the potential of conferring computational 
simplicity and yet avoiding the ambiguities in the boundary conditions 
on the throats in the calculations of GGB02.  The next section outlines 
the idea and its potential shortcomings.

\section{A combination of the conformal thin-sandwich formalism and the 
puncture method}
\label{sec:ctsp}

In order to apply the puncture method to the conformal thin-sandwich
decomposition, as a first step we need to examine how the lapse 
equation~(\ref{eqn:TSconstK}) and lapse function $\lapse$ behave in the
vicinity of the punctures.  From the similarities between the lapse equation 
(\ref{eqn:TSconstK}) and the Hamiltonian constraint (\ref{eqn:TSHCg}) we 
can anticipate splitting $\tilde{\lapse}\psi^7$ into a sum
\begin{equation}
\tilde{\lapse}\psi^7 
= 1 + \sum_n^N \frac{{\mathcal C}_n}{2 r_n} + v ,
\end{equation}
where the ${\mathcal C}_n$ are constants and $v$ is a correction.  The term 
$v$ is expected to be finite and sufficiently smooth that the domain once 
again can be taken as Euclidean.  As with the Hamiltonian constraint, for 
this to be true conditions must be imposed at the punctures on the behavior 
of $\tilde{A}^{ij}$, $K$, and now $\beta^i$.

The possible values of the ${\mathcal C}_n$ correspond to different boundary
conditions on the lapse at the other $N$ asymptotic infinities.  Each unique 
set $\{ {\mathcal C}_n \}$ gives rise to a different development between 
time slices.  The full generality of these boundary conditions need not be 
considered here.  It suffices for our present purpose to distinguish two 
principal classes.

To understand these two choices, consider a single Schwarzschild black 
hole described by isotropic spatial coordinates.  One might choose to 
describe the spacetime with static Schwarzschild time slices, for which 
the lapse function is
\begin{equation} 
\lapse = \frac{1 - M/2r}{1 + M/2r} .
\end{equation} 
Given that $\psi = 1 + M/2r$ in these coordinates, we have
\begin{equation} 
\lapse\psi = \tilde{\lapse} \psi^7 = 1 - \frac{M}{2 r} ,
\label{eqn:SSansatz}
\end{equation} 
which satisfies equation (\ref{eqn:TSconstK}) and the lapse serves to link 
successive maximal slices.  As is well known, these slices are frozen on 
the throat ($r=M/2$), have negative lapse on the second sheet, and 
$\lapse \rightarrow -1$ at the other spatial infinity.  In analogy 
with the arguments of Section \ref{sec:punc} this suggests that the 
generalization
\begin{equation} 
\tilde{\lapse} \psi^7 = 1 - \frac{1}{\psiana} + v 
\label{eqn:SchalphaSplit} 
\end{equation} 
be used for multiple black holes.  In this case the lapse would also 
be negative on each of the other $N$ sheets and limit on $-1$.  

Alternatively, one might choose to use the opposite sign on the singular
terms, to wit
\begin{equation} 
\tilde{\lapse} \psi^7 = 1 + \frac{1}{\psiana} + v .
\label{eqn:alphaSplit}
\end{equation} 
For a single Schwarzschild hole this reduces to 
$\tilde{\lapse}\psi^7 = 1 + M/2r$ and hence to $\lapse = 1$.  The unit lapse
does connect two slices in the foliation of maximal slices of Schwarzschild
found by Estabrook {\it et.~al.} \cite{estabrook73}.  Each pair of slices 
in that foliation has $\lapse > 0$ everywhere, though in general 
$\partial_i \lapse \ne 0$.  In applying this boundary condition to two or 
more black holes, we certainly expect $\partial_i \lapse \ne 0$ but, with a 
sufficient bound on $v$, will have $\lapse > 0$ and $\lapse \rightarrow 1$ 
at infinity on the other sheets.

These two choices illustrate the primary dichotomy in the choice of boundary
conditions on the lapse:  Either the lapse reverses sign in the interior of
the black holes, which in the puncture method means that each puncture is 
surrounded by a two-surface on which $\lapse = 0$, or the lapse remains 
positive everywhere, including on each {\sl then dynamic \/} throat.  In a 
nutshell, therein lie the potential shortcomings of combining the puncture 
method with the conformal thin-sandwich approach.

In the former case (\ref{eqn:SchalphaSplit}), the vanishing of the lapse 
on some surface around each puncture is very problematic for the solution 
of equation (\ref{eqn:TSMC}) or for the construction of the extrinsic 
curvature from equation (\ref{eqn:Aeqn}).  Examination of (\ref{eqn:TSMC})
reveals that $\tilde{\lapse} = 0$ is a critical surface for the differential
equations.  A regular solution for $\beta^i$ may be obtained by imposing 
the regularity condition that $m_j (\tilde{{\mathbb L}} \beta )^{ij} = 0$
on this surface, where $m_j$ is the normal to the level surface 
$\tilde{\lapse}=0$.  However, even if a regular solution for $\beta^i$ can 
be obtained, handling such an internal boundary is precisely the type of 
task that the puncture method seeks to avoid.  Furthermore, a regular 
solution for $\beta^i$ does not ensure a finite extrinsic curvature, as 
reference to equation (\ref{eqn:Aeqn}) indicates.  Stipulating the vanishing 
of all five algebraically-independent components of 
$(\tilde{{\mathbb L}}\beta )^{ij}$ on the critical surface appears to
entail more freedom than is available.  This is the difficulty that GGB02 
faced in requiring both isometry on the shift and regularity in 
$\tilde{A}^{ij}$.

So, if the use of the boundary conditions implied by equation 
(\ref{eqn:SchalphaSplit}) is doubtful, what of the latter choice expressed 
by equation (\ref{eqn:alphaSplit})?  Here the problem is more subtle.
As we discussed in Section {\ref{sec:equi}, it seems undesirable to 
construct a quasi-equilibrium solution through some means that leads to a 
dynamical slicing.  At minimum, this would make it difficult to verify 
whether or not one has in fact constructed a quasi-equilibrium solution.  
This requisite and the previous one lead to the question of whether it is 
possible to find a pair of slices (maximal or non-maximal) on which the 
fields appear stationary, that give rise to a lapse that is positive 
everywhere, and which end at spatial infinity on all sheets.  The second 
and third conditions are requirements on the applicability and usefulness 
of the puncture method.  The first condition attempts to ensure that the 
quasi-equilibrium nature of data constructed with the thin-sandwich formalism 
actually appear in equilibrium.  As we will argue below, such a slicing 
does not even exist for a Schwarzschild spacetime.

The argument is easy to make.  Consider a $t$ equal constant surface of 
Schwarzschild and $\lapse = 1$ consistent with (\ref{eqn:alphaSplit}).  
A quick evaluation of the evolution equation (\ref{eqn:Kdot}) indicates 
that the time derivative of $\tilde{A}^{ij}$ will be non-vanishing.  In
crossing the hole with a positive lapse we necessarily begin to see the
pinch-off of the throat.

This argument can be made more general by considering a standard Penrose
diagram of Schwarzschild, as in Figure~\ref{fig:Penrose1}.  The curved 
lines with arrows represent the Killing flow associated with the Killing 
vector that is timelike in the black hole exterior.  The Killing vectors 
are tangent to the $r= {\rm const}$ curves.  The two dashed
lines represent two spacelike hypersurfaces that have no special
properties other than ending at spatial infinity in both universes. In
particular, there is no reason that they should be maximal.  In this example, 
the upper surface represents a slice that stems from evolution off the 
lower slice with a lapse that is everywhere positive.

\begin{figure}[!ht]
\includegraphics[scale=0.4]{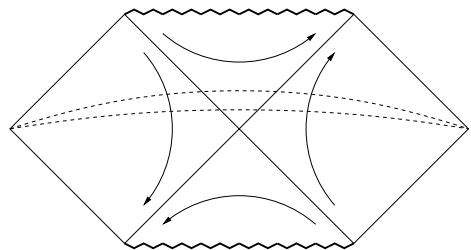} 
\caption{Penrose diagram of the Schwarzschild spacetime.  The curved solid 
lines are a Killing flow and the dashed lines are spacelike hypersurfaces.}
\label{fig:Penrose1}
\end{figure}

Figure~\ref{fig:Penrose2} focuses on the black hole interior and
just one of the hypersurfaces.  Here we begin with no {\it a priori \/} 
assumption about the second slice or whether the lapse changes sign or not
in crossing the black hole interior.  Point ``B'' represents the
minimal surface or throat in the hypersurface, as it corresponds to
the smallest value of the radial coordinate (e.g., areal coordinate).  
Any slicing that ends at spatial infinity in both universes must have 
such a point.

\begin{figure}[ht!]  
\includegraphics[scale=0.4]{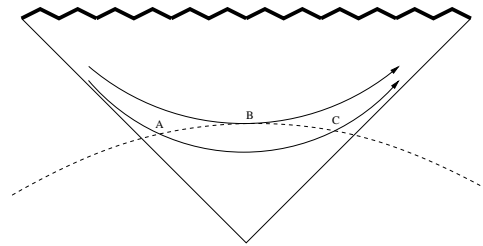} 
\caption{Penrose diagram of the black hole interior. Point B
represents the minimal two-surface.}
\label{fig:Penrose2}
\end{figure}

This leads to the following crucial observation.  As we are interested in 
stationary slicings, the time four-vector $t^{\mu}$ ($\partial/\partial t$) 
for evolving to the next slice should be coincident with the Killing vector.
The spatial slice is tangent to the $r= {\rm const}$ surface at the
throat.  But the Killing vector is tangent to the $r= {\rm const}$
surface everywhere.  So, in particular, the Killing vector is tangent
to the spatial slice at the throat.  The time four-vector can be written
in terms of the lapse and shift as $t^{\mu} = \lapse n^{\mu} +
\beta^{\mu}$.  At the location in question, since the Killing vector is 
tangent to the hypersurface, the lapse must vanish at the throat.
This is unavoidable and establishes that we cannot choose a slicing that
connects two spatial infinities, is time independent, and has an
everywhere positive lapse.  The lapse must at least vanish at the
minimal surface if the slicing is stationary.  

The argument can be extended to show that if stationarity is assumed the 
lapse must become negative interior to the throat.  At point ``C''
in the figure the Killing vector points in the upper right direction so the
lapse must be positive there.  But, at point ``A'', the Killing vector
points in the lower right direction, so the lapse will be negative at ``A''.
Alternatively, if the lapse is positive definite, the slice must be 
non-stationary at least in some neighborhood of the black hole throat.

An added consequence of these arguments is that a slicing of Schwarzschild 
that is stationary and has an everywhere positive lapse can never be tangent 
to an $r= {\rm const}$ surface.  This requires that one end of the slice 
terminate at the physical singularity, which obviates the application of 
the puncture method.

\section{Summary}
\label{sec:summ}

Several arguments point to the conformal thin-sandwich decomposition
as a promising approach for constructing binary black holes in
quasi-circular orbits (cf. GGB02).  However, assuming an isometry and
using the black hole throats as inner boundaries leads to some
mathematical inconsistencies (cf. Ref.~\cite{cook02}).  This
experience suggested that one might combine the conformal
thin-sandwich approach with the puncture method \cite{brandt97}, which
seeks to avoid interior boundaries altogether.

Combining the conformal thin-sandwich decomposition with the puncture
method for the construction of equilibrium data imposes three
requirements on the slicing.  Equilibrium data can most easily be
identified when they are presented in stationary slices.  The puncture
method also requires that the slices connect the spatial infinities of
the multiple black holes, as opposed to ending on a physical singularity.  
Finally, the conformal thin-sandwich decomposition appears to require an 
everywhere positive lapse, to ensure a finite extrinsic curvature and, at 
minimum, to avoid internal boundaries.  However, even a Schwarzschild 
spacetime admits no stationary slicing that has an everywhere positive 
lapse and ends at spatial infinity in both universes.

The above arguments do not mean that it is impossible to construct
quasi-equilibrium data with a combination of the conformal thin-sandwich 
approach and the puncture method.  One might resign to constructing data 
in a slicing that cannot be everywhere stationary.  For example, a 
construction might yield slicings that are quasi-stationary exterior to 
a neighborhood of the throat but become dynamical near the throat.  The
issue then is establishing whether the resulting data are in fact in 
quasi-equilibrium.  Not enough is known yet to rule out the success of this
technique.  Alternatively one might consider allowing the lapse to cross 
zero.  It is not yet completely clear what regularity conditions can be 
imposed or how the extrinsic curvature behaves in the neighborhood of these
surfaces.  At minimum this undertaking potentially involves a difficult 
numerical implementation at the critical surfaces; the need to do so 
defeating the purpose of the puncture method.  We conclude that the use of
a combination of the conformal thin-sandwich and puncture methods for 
constructing quasi-equilibrium black hole initial data is less 
straightforward than it first appeared and may in fact be impractical. 

\acknowledgments

We would like to thank Niall \'O~Murchadha and Bernd Br\"ugmann for
illuminating discussions.  CRE and GBC thank the Kavli Institute for
Theoretical Physics (KITP) where a portion of this work was completed.
This research was supported in part by the National Science Foundation
under grants PHY-9907949 (to KITP), PHY-0140100 (to Wake Forest
University), PHY-0207097 (to the University of North Carolina), and
PHY-0139907 (to Bowdoin College).

\bibliography{ctspunc}

\end{document}